\documentclass[
showpacs,floatfix, aps,prl,amsmath,amssymb,
twocolumn,
groupaddress
]{revtex4}

\def\XXint#1#2#3{{\setbox0=\hbox{$#1{#2#3}{\int}$}
     \vcenter{\hbox{$#2#3$}}\kern-.5\wd0}}

\usepackage{graphicx}
\usepackage{feynmp}

\renewcommand{\r}{{\boldsymbol{r}}}

\newcommand{\nnabla}{{\boldsymbol{\nabla}}}
\newcommand{\p}{{\boldsymbol{p}}}

\newcommand{\be}{\begin{equation}}
\newcommand{\ee}{\end{equation}}
\newcommand{\bea}{\begin{eqnarray}}
\newcommand{\eea}{\end{eqnarray}}
\newcommand{\bs}{\begin{split}}
\newcommand{\bes}{\begin{equation}\begin{split}}
\newcommand{\ees}{\end{split} \end{equation}}
\newcommand{\es}{\end{split}}

\newcommand{\D}{\mathcal{D}}

\newcommand{\G}{\mathcal{G}}

\newcommand{\vc}[1]{\mathbf{#1}}

\begin{document}

\title{Bound State of Ultracold Fermions in the Spin Imbalanced Normal State}

\author{F. Fumarola, I.L. Aleiner and B.L. Altshuler}
\affiliation{Physics Department, Columbia University, New York, NY 10027}

\pacs{03.75.Ss
,05.30.Fk
,03.75.Kk
}
\begin{abstract}
We study a polarized Fermi gas and demonstrate that Fano-Feshbach
(FF) resonances lead to the pairing of fermions and holes into long
living massive bosonic modes (bifermions and biholes), which can be
viewed as signatures of a first order quantum phase transition.
These modes, which are neither conventional Cooper pairs nor closed
channel bosonic dimers, have a peculiar dispersion relation and
appear already in the weak coupling limit, i.e. for large FF
detunings. Our result is in accord with claims of Schunck \emph{et
al.} [cond-mat/0702066] that by stabilizing the normal state the
population imbalance still allows effect of s-wave fermionic
pairing.
\end{abstract}
\date{\today}
\maketitle

Quantum phase transitions in polarized Fermi gases have been a
subject of investigation for many years \cite{ClCh,
fflo1,fflo2,imbalance,leo}. In Ref. \cite{ClCh}, it was demonstrated
that spin polarization of electrons in a superconductor leads to a
first order quantum phase transition. An applied magnetic field
induces a Zeeman splitting $E_Z$ in the energies of spin up and spin
down electrons and imbalances the equilibrium densities of the two
species. Eventually, for large enough $E_Z$ the superconducting
ground state becomes a polarized normal metal rather than a
superconductor.

Let us define the imbalance in the system as $J =\frac{ N_\uparrow -
N_\downarrow}{N_\uparrow + N_\downarrow}$, $N_\uparrow$ and
$N_\downarrow$ being the number of spin up and spin down fermions.
To form an s-wave Cooper pair, two fermions have to be in different
spin states. Therefore, for $J=1$ the system has normal ground
state. On the other hand, the ground state of a Fermi gas with no
population imbalance ($J=0$) is superfluid. Since no crossover
between two states of different symmetry is possible, the system
will remain normal even at zero temperature ($T=0$) as long as $J$
exceeds certain critical value $J_c$.

In superconductors, this translates into the existence of a critical
magnetic field: in particular, the phase diagram in the presence of
Zeeman (acting only on the spins) magnetic field \cite{impur}
includes the BCS phase for $E_Z < \Delta$, a cohesistence region
$\Delta < E_Z < \sqrt{2} \Delta$, and the normal state for $E_Z >
\sqrt{2} \Delta$, where $\Delta$ is the BCS superconducting gap at
zero temperature in the absence of Zeeman splitting. The elusive
Fulde-Ferrell-Larkin-Ovchinnikov (FFLO) state \cite{fflo1,fflo2}
appears in a narrow region below $E_Z= 1.508 \Delta$: in this state,
the mismatch of the spin up and spin down Fermi surfaces
 induces the electrons to form pairs with finite momentum.
As a result, the superconducting gap becomes a periodic function of
the space variables.

Experimental studies of superfluidity have been performed in
ultracold Fermi gases \cite{biglistE,chin} in the BEC-BCS crossover,
tuning the scattering length through Fano-Feshbach (FF) resonances
\cite{FanoFeshbach,review}. Effect of the polarization was achieved
by creating imbalanced mixtures of ultracold $^6$Li atoms in two
different hyperfine states (which we will refer to as spin states
$\uparrow, \downarrow$). In the presence of the imbalance, anomalous
density profiles \cite{hulet} and the disappearance of quantum
vortices \cite{ketterle} have been reported. While in metals orbital
effects associated with the magnetic field can obscure the effects
of Zeeman splitting, in neutral atomic gases this problem does not
exist. Moreover, while in metals the relaxation due to spin orbit
interaction tends to equilibrate the spin populations, in ultracold
atomic gases the spin is a good quantum number, and the initial
imbalance in the system is conserved.  In these systems, the
difference between the numbers $N_\uparrow$ and $N_\downarrow$ of
majority and minority atoms can be encoded by the difference $2 h$
in the chemical potential of the two spin states, which plays the
same role as the Zeeman splitting for electrons in metals.

In fact, the transition from superfluid to normal state in polarized
Fermi gases (unlike the transition in an unpolarized gas) is an
example of first order quantum phase transitions. Being
discontinuous, first order quantum phase transitions appear to have
no critical manifestations. However, as it was pointed out in
\cite{igor}, the signatures of such phase transitions are the
anomalies (massive modes) in the \emph{excitation} spectrum. These
modes lead to sharp features in the observables and can determine
the collective quantum dynamics.
 Moreover, the study of such modes may provide valuable information
about the relaxation processes.

\begin{figure}[!tb]
\!\!\!\!\!\!\!\!\vspace{.1cm}
\includegraphics[height=2.5in,width=3.5in,angle=0]{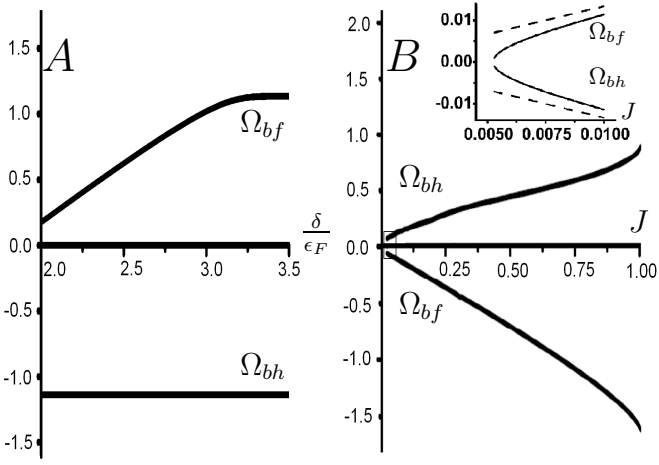}
\put(-16,132){$J$} \put(-28,162){$\Omega_{bf}$}
\put(-28,145){$\Omega_{bh}$}
\begin{large}\put(-160,127){$\Omega_{bf}$}
\put(-160,35){$\Omega_{bh}$} \put(-100,58){$\Omega_{bf}$}
\put(-100,106){$\Omega_{bh}$} \put(-13,89){$J$}
\put(-137,86){$\frac{\delta}{\epsilon_F}$}\end{large}
\begin{LARGE}\put(-235,150){$A$}\put(-105,150){$B$}
\end{LARGE}\caption{Excitation gaps $\Omega_{bf}(q=0)$ and
$\Omega_{bh}(q=0)$ of bifermions and biholes in units of Fermi
energy as functions of detuning $\delta$ for fixed imbalance
$J=\frac{ N_\uparrow - N_\downarrow}{N_\uparrow + N_\downarrow}=0.8$
 (panel A) and as functions of the
imbalance $J$ for $\delta=2.5 \epsilon_F$ (panel B).  The inset
shows the behavior of the gaps for small imbalance, in the same
units, together with the boundaries of the continuum (dashed lines).
Here, the resonance width is $\gamma=0.1$. The symmetry in the
spectrum of
 bifermions
and biholes ($\Omega_{bf} \sim - \Omega_{hf}$) is recovered for
large detuning in panel A and for small imbalances in panel B.
}
 \label{graph1}
\end{figure}

The identification of the massive modes and evaluation of their
spectrum is a non-trivial model dependent task. The purpose of this
Letter is to carry out this program for the normal state of
resonantly coupled fermions with population imbalance. We prove that
arbitrarily narrow FF resonances lead to the appearance of massive
modes if $J$ exceeds the critical value $J_c$ so that the normal
state is stable. We identify two kinds of bosonic quasiparticles:
bifermions and biholes, having spin zero, associated with the
pairing of fermions and holes. These modes have different features
both from the closed channel dimers of the BEC-BCS crossover and
from conventional Cooper pairs. They have a strongly $J$-dependent
gap (shown in Fig. \ref{graph1}), and a peculiar dispersion
relation, their energy being a decreasing function of momenta.

The FF resonant Fermi gas can be modeled by the following
grandcanonical hamiltonian:
\bea \hat{H}&=&\int d\r \left[ \sum_\sigma \hat{f}^\dagger_\sigma
\epsilon^f_\sigma\left(-i\hbar\nnabla\right)\hat{f}_\sigma +
\hat{b}^\dagger\epsilon^b\left(-i\hbar\nnabla\right)\hat{b} \right]
\nonumber
\\
&+&g \ \mathrm{H.p} \int d\r \sum_{\sigma_1\sigma_2}
\left[\sigma^y_{\sigma_1\sigma_2}\hat{b}
\hat{f}^\dagger_{\sigma_1}\hat{f}^\dagger_{\sigma_2}\right],
\label{model} \eea \vspace{-.5cm} \be
\epsilon^f_\sigma=\frac{\p^2}{2m}  - \mu + 4 \sigma h; \quad
\epsilon^b= \frac{\p^2}{4m} - 2 \mu + \delta_0, \nonumber \ee
 where $\sigma^y$ is a Pauli matrix, H.p. denotes the
Hermitian part of an operator,
$\hat{f}^\dagger_\sigma(\r),[\hat{f}_\sigma(\r)]$ create
[annihilate] fermionic atoms with mass $m$ and spin $\sigma=\pm
1/2$, while $\hat{b}^\dagger(\r),[\hat{b}(\r)]$ create [annihilate]
bosonic molecules; $\mu$ is the chemical potential. The bare
detuning $\delta_0$ depends on the high momentum cutoff $p_0
\rightarrow \infty$. It is related to the position of the FF
resonance $\delta$, which is independent of $p_0$, and is connected
with the s-wave scattering length $a_s$ by:
\bea \label{eq:deltaintro} \delta_0 = \delta + g^2 \int^{p_0}
\frac{d^3 p}{(2\pi\hbar)^3} \frac{m}{p^2} \qquad  a_s= - \frac{m
g^2}{4 \pi \hbar \delta}. \eea

\begin{figure}[!tb]
\includegraphics[height=2.3in,width=3in,angle=0]{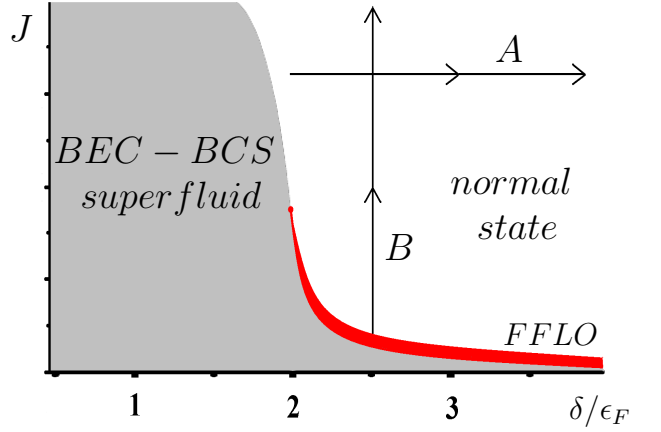}
\begin{Large}\put(-85,65){$B$} \put(-43,140){$A$}
 \end{Large} \begin{large} \put(-40,33){$FFLO$} \end{large}
\begin{large}\put(-15,5){$\delta/\epsilon_F$} \end{large}
\begin{Large} \put(-227,148){$J$} \end{Large} \begin{Large}
\put(-210,102){$BEC-BCS$} \put(-200,85){$superfluid$}
\end{Large} \begin{Large} \put(-60,90){$normal$} \put(-50,73){$state$}
 \end{Large} \caption{ (Color online) Schematic picture of the phase diagram at
small coupling (for a complete discussion, see \cite{leo}). Arrows
indicate the trajectories corresponding to the plots in Fig.
\ref{graph1}. The FFLO phase lies in the red region at the
superfluid/normal state boundary.} \label{phasediagram}
\end{figure}

The key dimensionless parameter is $\gamma =\frac{ \sqrt{8} g^2
m^{3/2}}{ 4\pi^2 \hbar^3 \sqrt{\epsilon_F}}$, the ratio of the width
of the FF resonance to the Fermi energy $\epsilon_F$ ($\epsilon_F
\equiv \mu$ at $J=0$). The model
(\ref{model}) 
suggests that in the weak coupling limit ($\gamma \ll 1$) that the
system would be in the normal phase as long as  $\delta >
\delta_c(J)\sim 2 \epsilon_F$ (Fig. \ref{phasediagram}). However,
current experiments performed with wide resonances ($\gamma \gg 1$)
have shown that the system remains in the normal state down to much
smaller values of the detuning: at resonance ($\delta =0$), $J_c$
was found as small as $0.7$ \cite{ketterle}. We believe that our
conclusions remain qualitatively valid for wide resonances and at
resonance, as long as the normal state is stable.

\begin{figure}[!tb]
\includegraphics[height=2.4in,width=3.5in,angle=0]{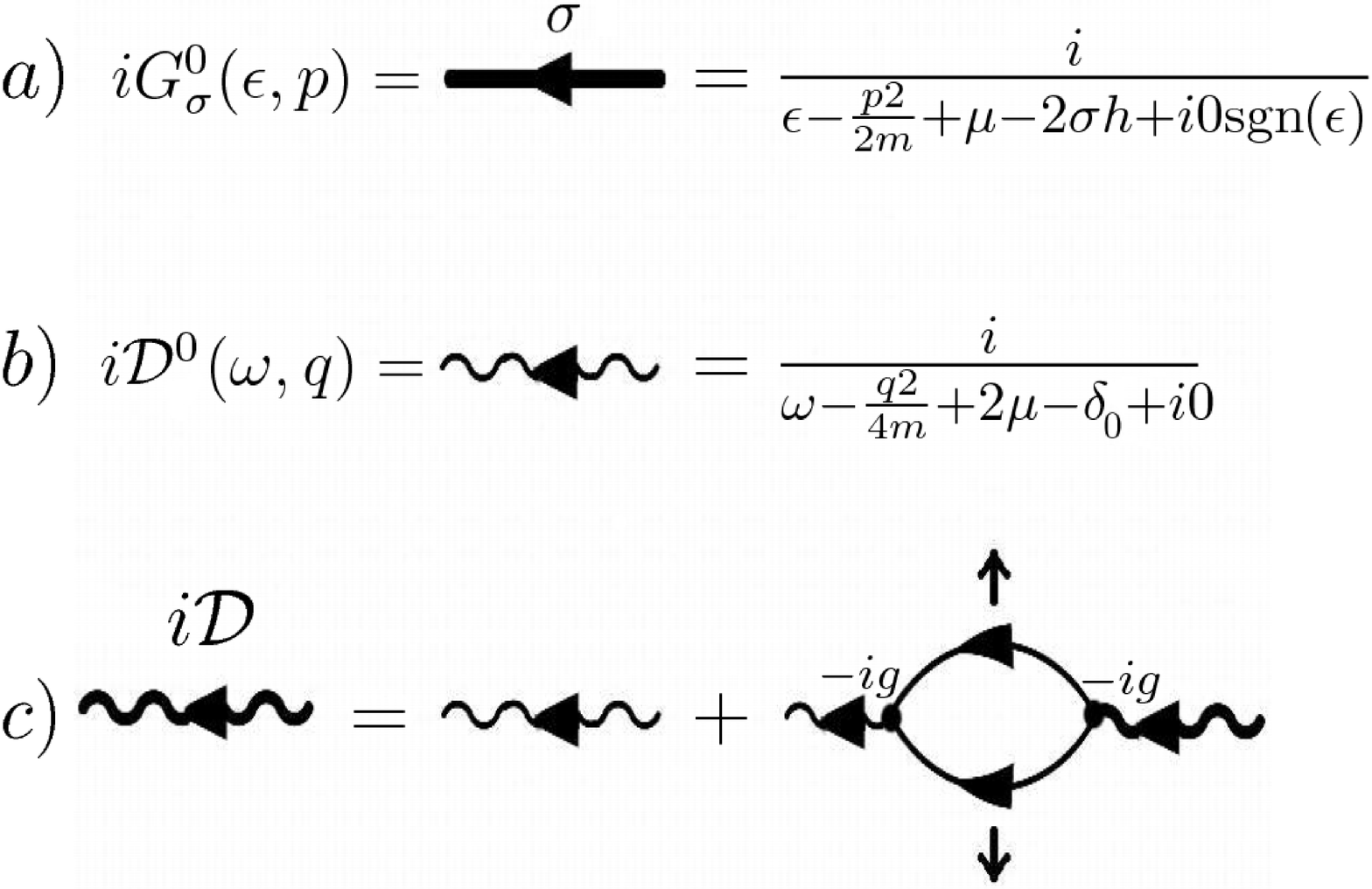}
\caption{Diagrammatic representation of the Dyson equation for the
particle-particle propagator, in the zero temperature formalism.}
\label{stk}
\end{figure}

In the case of narrow resonances ($\gamma \ll 1$), we can employ the
one loop approximation for the bosonic Green function $\D(x_1,x_2)$,
which amounts to the Dyson equation in Fig. \ref{stk}, $\D^0$ and
$\G^0$ being the non-interacting Green functions for the bosons and
fermions correspondingly.

Since $\D$ is nothing but the two fermion scattering amplitude,
bound states and resonances of the system are found by identifying
its poles. Solving Eq. c) of Fig. \ref{stk} by means of Eqs. a) and
b), after integration over frequencies, we find the equation for the
dispersion relation $\Omega(q)$ of excitations in the
particle-particle channel at zero temperature
\bea &&  \!\!\! \sum_{\pm} \int_0^\infty \!\!\!\!\!\!
\sqrt{\epsilon} d \epsilon\!\! \int_{-1}^1 \!\!\frac{d x}{4} \left[
- \frac{1}{ 2 \epsilon}\! + \frac{\mathrm{sgn}(\epsilon +
\frac{q^2}{8 m} - \mu \pm h + q \sqrt{\frac{\epsilon}{2 m}} x) }{2
\epsilon + \frac{q^2}{4 m} - 2 \mu - \Omega(q)} \right]   \nonumber \\
&& + \ \frac{\Omega(q) - \frac{q^2}{4 m} + 2 \mu - \delta }{\gamma
\sqrt{\epsilon_F}}=0. \label{spettro} \eea
Here, $h$ and the chemical potential $\mu$ are the Lagrange
multipliers corresponding to the conserved quantities $N_{\uparrow}
- N_{\downarrow}$ and $N_{\uparrow} + N_{\downarrow}$. In our
discussion, $h$ and $\mu$ can be calculated by setting $\gamma
\rightarrow 0$, since any correction to the zero coupling limit
would translate into corrections to Eq. (\ref{spettro}) of higher
order in the small parameter $\gamma \ll 1$. Hence, given the
imbalance $J$ and the density of atoms $n$, one can determine $h$
and $\mu$ from
\be \begin{split} \label{legendre} & \frac{\sqrt{2}}{3 \pi^2}
m^{3/2} (\mu
+ h)^{3/2} + \frac{\sqrt{2}}{3 \pi^2} m^{3/2} (\mu - h)^{3/2} = n \\
& \frac{\sqrt{2}}{3 \pi^2} m^{3/2} (\mu + h)^{3/2} -
\frac{\sqrt{2}}{3 \pi^2} m^{3/2} (\mu - h)^{3/2} = J n. \end{split}
\ee

Equation (\ref{spettro}) has two solutions $\Omega_{bf}(q)$ and
$\Omega_{bh}(q)$ which correspond to the two kinds of
quasiparticles. These excitations have gaps, which can be found from
the equation for $\Omega$

\be \begin{split} & \sqrt{\mu + \frac{\Omega}{2}}\left(
\mathrm{tanh}^{-1}\left[\frac{\sqrt{\mu - h}}{\sqrt{\mu +
\Omega/2}}\right] + \mathrm{coth}^{-1}\left[\frac{\sqrt{\mu +
h}}{\sqrt{\mu + \Omega/2}}\right]\right) \\ & - \sqrt{\mu + h}
-\sqrt{\mu - h} + \frac{\Omega + 2 \mu - \delta}{\gamma
\sqrt{\epsilon_F}} = 0
\end{split} \ee

The dependence of the gaps $\Omega_{bf}$ and $\Omega_{bh}$ on $J$
and $\delta$ is plotted in Fig. \ref{graph1}.

Clearly, these quasiparticles have spin zero, and obey bosonic
statistics. However, they are quite different from conventional
Cooper pairs in that their energy is a decreasing function of
momenta and their excitation gap strongly depends on the imbalance
$J$ of the system. Bifermions and biholes are also quite different
from the closed channel bosonic dimers of the BEC-BCS crossover: in
the limit $\delta \gg 2 \epsilon_F$, such dimers are too energetic
to play any role in the low energy physics of the Fermi gas. In
contrast, bifermions and biholes can be observed experimentally even
at large $\delta$. Their excitation gap $\Omega_0 \sim 2 h$ can be
by orders of magnitude smaller than the binding energy $\delta$ of
the bare bosonic dimers.

In the limit $J \ll 1$, the quasiparticle spectrum can be evaluated
analytically. Equation (\ref{spettro}) becomes
\be \!\!\! \sum_{\sigma, \tau = \pm 1} \!\! \frac{2 \sigma h + v_F q
+ \sigma \tau \Omega(q)}{v_F q}\ln\left[\frac{2 h + \sigma v_F q +
\tau \Omega(q)}{\Delta}\right] = 4 \nonumber \ee 
with $h = \frac{2}{3} J \epsilon_F$. Here,
 $\Delta$ is the $T=0$ superfluid
order parameter, which satisfies the gap equation

 \be \label{gappetto} \frac{\delta -
2\mu}{\gamma \epsilon_F} = \frac{1}{2} \int_{-\mu}^{\infty} d\xi
\sqrt{1+\frac{\xi}{\mu}} \left[ \frac{1}{\sqrt{\xi^2 + \Delta^2}}
-\frac{1}{\xi + \mu}\right] \ee
For $\delta \gg 2 \epsilon_F$, $\Delta \sim 8 \epsilon_F \exp[-2 -
\frac{\delta/\epsilon_F - 2 }{\gamma}]$.

The dispersion relation of these quasiparticles is nontrivial. For
$J \ll 1$, bifermions and biholes turn out to have opposite spectra.
In this limit, as well as in the limit $\delta \gg \epsilon_F$ (see
Fig. \ref{graph1}), bifermions and biholes can be viewed as
particles and antiparticles. For $q \ll J \sqrt{m \epsilon_F}$, the
dispersion relation is given by:
\be \begin{split} \label{main} & \Omega_{bf}(q) = - \Omega_{bh} (q)=
\Omega_{bf}(0) - \frac{1 + 2 (\Omega_{bf}(0) / \Delta)^2 }{6 \Omega_{bf}(0)} v_F^2 q^2\\
& \Omega_{bf}(q=0) = \sqrt{\left( \frac{4}{3} J \epsilon_F \right)^2
- \Delta^2}  \end{split} \ee
where $v_F = \sqrt{2 \epsilon_F/ m}$. According to Eq. (\ref{main}),
which is our main analytic result, the excitation energy is a
decreasing function of momenta, and the excitation gap depends on
the coupling between the fermions through the superfluid order
parameter $\Delta$. 

\begin{figure}[!tb]
\includegraphics[height=3in,width=3.3in,angle=0]{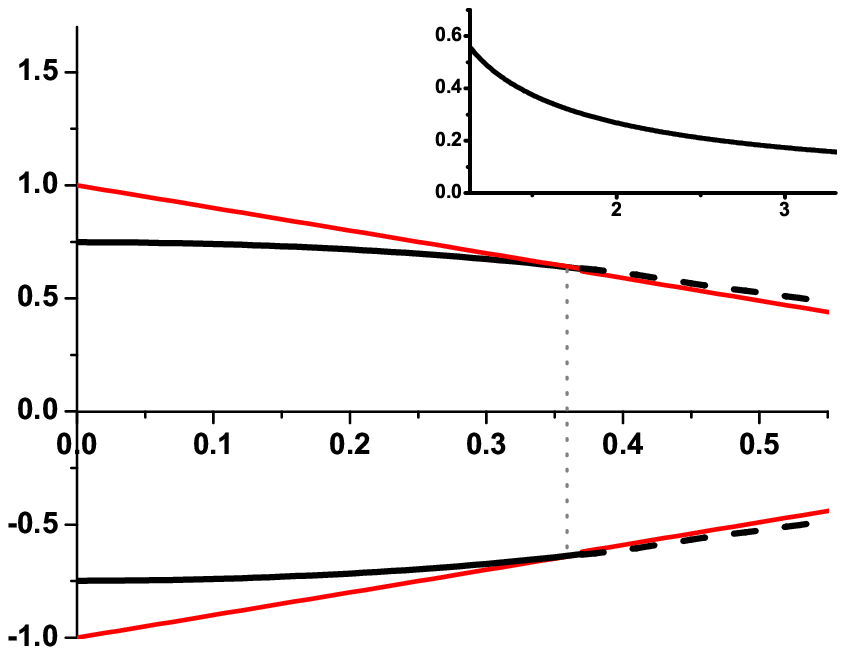}
\put(-20,140){$J \epsilon_F / \Delta$} \put(-125,190){$q_c$}
\begin{large} \put(-190,119){$\Omega_{bf}$}\put(-190,50){$\Omega_{bh}$}
\put(-15,70){$q$}\put(-90,77){$q_c$}
\end{large}
\caption{(Color online) The main plot shows the dispersion relations
of bifermions and biholes,  for $J = J_c$ (frequencies are in units of $2 h$, momenta in units of $2 h / v_F$). 
 The value $q_c$
for which the spectrum intersects the cuts $\Omega= \pm (2h - v_F
q)$ (red lines in the main plot) is plotted in the inset as a
function of $J \epsilon_F/\Delta$, in units of $\Delta /v_F$. }
 \label{lowmom}
\end{figure}

In the one loop approximation, bifermions and biholes are undamped
up to the critical value of momentum $q_c$ \cite{footnote}, where
their spectrum intersects a cut of the bosonic Green function (Fig.
\ref{lowmom}). For $q>q_c$, the collective modes develop a nonzero
imaginary part.

The vanishing of $\Omega_{bf}(q=0)$ and $\Omega_{bh}(q=0)$ at $J=
\frac{3}{4} \frac{\Delta}{\epsilon_F}$ manifests the instability of
the normal state with respect to formation of the homogeneous
$\Delta$ for $J< \frac{3}{4} \frac{\Delta}{\epsilon_F}$. This
corresponds to the first order phase transition mentioned above. As
to the instability toward inhomogeneous $\Delta(\vc{q})\sim e^{i
\vc{q} \vc{r}}$, it appears at $\tilde{J}_c = 1.13
\frac{\Delta}{\epsilon_F}$ \cite{fflo2}, corresponding to a second
order quantum phase transition to the FFLO phase with broken
translational symmetry.

The critical behavior near the second order transition between the
normal and FFLO state at $J<J_c$, as well as the collective
excitations, definitely deserve theoretical analysis. Here we can
note just that the bifermion and bihole modes have little to do with
these problems. Namely, the analysis of the solution of Eq.
(\ref{spettro}) on the physical and nonphysical sheets suggests that
the FFLO transition manifests itself in the appearance of new
diffusive branch rather than in softening of the mode  (\ref{main}).

In the context of disordered superconductors, the existence of
massive modes was discovered through measurements of the tunneling
density of states \cite{adams}, and the theory of the effect was
constructed in Ref. \cite{igor}. We believe that bifermions and
biholes in ultracold atomic gases should manifest themselves in
measurements of radio-frequency spectroscopy, which have detected
superfluidity at $J=0$ \cite{chin}. Possible additional experimental
applications of those modes will be discussed elsewhere \cite{us}.

In conclusion, we have shown that FF resonances lead to collective
modes in a Fermi gas with spin imbalance. As long as the imbalance
is large enough to prevent the formation of inhomogeneous
configurations, the physics is dominated by the appearance of
bosonic quasiparticles (bifermions and biholes). These
quasiparticles have finite gap, anomalous
 dispersion relation, and exist as long as the system is in the
normal state.

\emph{Added Note}. After this paper was completed, experimental data
were published by Shunck \emph{et al.} \cite{ketterlenew},
suggesting that s-wave pairing of atoms indeed takes place in the
spin imbalanced normal phase of ultracold fermionic gases, as an
explanation of anomalies in the population of imbalanced $^6Li$
atoms after applied radiofrequency. We find it likely that bifermion
modes, which we discussed here provide a coherent interpretation of
those anomalies.


\begin{thebibliography}{}
\bibitem{impur} The orbital effect of the magnetic field can be made
negligeable in superconductors in the presence of magnetic
impurities or in restricted geometries.
\bibitem{footnote} Higher loops corrections would provide a finite
imaginary part to the spectrum of bifermions and biholes, because
the decay into four fermions is not forbidden by momentum
conservation even for $q=0$.
\end{thebibliography}
\end{document}